\documentclass[12pt]{iopart}

\newcommand{\be}{\begin{equation}}
\newcommand{\ee}{\end{equation}}
\newcommand{\nn}{\nonumber}
\usepackage{iopams}  
\usepackage{mathrsfs}

\begin{document}

\title[Black Holes and Scalar Fields]{Black Holes and Scalar Fields}

\author{Thomas P.~Sotiriou}

\address{School of Mathematical Sciences \& School of Physics and Astronomy, University of Nottingham, University Park, Nottingham, NG7 2RD, UK}
\address{Perimeter Institute for Theoretical Physics, Waterloo, Ontatio, N2L 2Y5, Canada}
\ead{Thomas.Sotiriou@nottingham.ac.uk}
\vspace{10pt}
\begin{indented}
\item[]February 2014
\end{indented}

\begin{abstract}
No-hair theorems in theories of gravity with a scalar field are briefly and critically reviewed. Their significance and limitations are discussed and potential evasions are considered.\footnote{Contribution to the Classical and Quantum Gravity Focus Issue ``Black holes 
and fundamental fields''.}

\end{abstract}

%
%
%
%
%

\section{Introduction}

There are at least 3 related but distinct reasons for which one might be interested in the behaviour of scalar fields in black hole spacetimes:

In vacuum general relativity and in Einstein--Maxwell theory there are black hole uniqueness theorems that pin down the Kerr--Newman 3-parameter family of solutions as the only admissible one \cite{Hawking:1971vc,israel1, israel2, Carter:1971zc, Wald:1971iw}. Two of these parameters are the mass and the angular momentum and the third is an electromagnetic charge. It has  been  conjectured that black holes cannot carry any other charge, and this is the so-called no-hair conjecture \cite{wheeler}.  In order to check if this conjecture is true and understand the structure of black holes one needs to explore different field contents. Scalar fields is one of the interesting cases and the one we will focus on here.  In passing, it is worth mentioning that hairy black hole solutions do exist in Einstein--Yang--Mills theory (see Ref.~\cite{Volkov:1998cc} for a review).

The second piece of motivation comes from the need to constrain alternative theories of gravity. Many alternatives to general relativity contain scalar fields (see Ref.~\cite{Sotiriou:2015lxa} for a brief review). If the existence of such scalars leads to different black hole spacetimes than those of general relativity, or simply new phenomena associated with the presence of the scalar field as we will see below, one could hope to detect these deviations in astrophysical observations.

Finally, one might be interested in the detection of scalar fields themselves, and black holes could just be the right laboratory for doing so. 

With this 3-fold motivation in mind, I will discuss in what follows the role of scalar fields in the structure of black holes in different theories of gravity. As a warm up, the next section focuses on scalar field configurations in curved background. This simplified analysis is actually the basis of the well-known no-hair theorems for general relativity and scalar-tensor theories. Section \ref{ST} focuses on scalar-tensor theories and contains a review of no-hair theorems, as well as a critical discussion about their significance and astrophysical relevance. Section \ref{othertheories} is devoted to generalised scalar-tensor theories, namely Horndeski theory and theories where a scalar field defines a preferred foliation. Some characteristic solutions with scalar hair are reviewed and the restrictions under which one can formulate no-hair theorems in Horndeski theories are discussed. Section \ref{concl} contains conclusions.

This is not meant to be an exhaustive review of scalar field phenomenology in black hole spacetimes. Even though I will briefly touch upon some interesting effects associated with scalar fields in black hole spacetimes, such us superradiance,  spontaneous scalarization, floating orbits, hairy black holes, etc., most of these topics will be covered in more depth in other contributions to this special issue. The main focus of this contribution will be the role of the scalar field in the structure of the black hole itself.

\section{Scalar fields on a black hole background}
\label{theorems}

Consider a scalar field propagating on a black hole background. Its equation of motion is
\be
\label{wave}
\Box \phi=0\,,
\ee
where $\Box\equiv g^{\mu\nu}\nabla_\mu\nabla_\nu$, $g^{\mu\nu}$ is the inverse of the spacetime metric, and $\nabla_\mu$ is the associated covariant derivative. Assume now that the background is
\begin{itemize}
\item stationary, as the endpoint of gravitational collapse;
\item asymptotically flat, as an isolated object.
\end{itemize}
Stationarity implies that there exists a Killing vector, $\xi$, that is timelike at infinity. If the spacetime is to be a vacuum solution of Einstein's equations (for the moment the scalar is assumed to be a ``probe field" propagating in this spacetime without back-reaction), then it will also have to be axisymmetric, by virtue of a theorem by Hawking in Ref.~\cite{Hawking:1971vc}. This means that there ought to be a second Killing vector, $\zeta$, that has closed orbits. One could make axisymmetry an extra assumption, in which case one would not need to assume that the metric satisfies vacuum Einstein equations.

Assume now that the scalar field respects the symmetries of the metric, {\em i.e.}~the Lie derivatives of the scalar with respect to the Killing vectors, $\mathcal{L}_\xi \phi$ and $\mathcal{L}_
\zeta \phi$ both vanish. This condition can also be written as $\xi^\mu \nabla_\mu \phi=\zeta^\mu \nabla_\mu \phi=0$. Assume also that $\phi\to \phi_0$ at infinity, where $\phi_0$ is some constant. Now consider the volume ${\cal V}$ bounded by a timelike 
3-surface at infinity, part of the black hole horizon, and two surfaces, ${\cal S}_1$ and ${\cal S}_2$. ${\cal S}_1$ is a partial Cauchy hypersurface for 
$\bar{J}^+( \mathscr{I}^-)\cap\bar{J}^-( \mathscr{I}^+)$, {\em i.e.}~the 
intersection of the topological closure of the causal future of 
past null infinity with the topological closure of the causal 
past of future null infinity, and ${\cal S}_2$ is obtained from ${\cal S}_1$ by shifting each point of ${\cal S}_1$ by a 
unit parameter distance along integral curves of $\xi^\mu$. Multiplying both sides of eq.~(\ref{wave}) by $\phi$ and integrating over ${\cal V}$ yields
\be
\int_{\cal V} d^4 x\sqrt{-g}\, \phi\Box \phi= 0\,,
\ee
where $g$ is the determinant of the metric. Integrating by parts, one obtains
\be
\int_{\cal V} d^4 x\sqrt{-g}\, \nabla^\mu \phi\nabla_\mu \phi =\int_{\partial{\cal V}} d^3 x\sqrt{|h|}\, n^\mu \nabla_\mu\phi \,,
\ee
where $\partial{\cal V}$ denotes the boundary of ${\cal V}$, 
$n^{\mu}$ is the normal to the boundary, and $h$ is the 
determinant of the induced metric, $h_{\mu\nu}$, on the 
boundary. It is rather straightforward to argue that the boundary contribution on the right hand side vanishes. The contribution from the timelike 3-surface at infinity will vanish due to asymptotic condition $\phi\to \phi_0$. The contribution from the horizon will also vanish, as on the horizon $n^\mu$ is a linear combination of $\xi$ and $\zeta$. The contributions from the two partial Cauchy surfaces will not vanish, but they will precisely cancel each other. One then has 
\be
\int_{\cal V} d^4 x\sqrt{-g}\, \nabla^\mu \phi\nabla_\mu \phi =0 \,.
\ee
However, the gradient of the scalar cannot be timelike anywhere or null everywhere as it is orthogonal to both of the Killing vectors. Therefore, the above integral can only vanish if $\partial_\mu \phi=0$ everywhere. Taking into account the asymptotic condition yields $\phi=\phi_0$ as the unique solution. 

This proves that the only admissible scalar configuration in a stationary, axissymetric, asymptotically flat black hole back-ground is a trivial one, provided that the scalar respects the symmetries of the metric and has trivial asymptotics. The proof has been first presented in Ref.~\cite{Hawking:1972qk}.

A straightforward generalisation of the proof that includes a potential for the scalar has been presented in Ref.~\cite{Sotiriou:2011dz}. One starts with the equation
\be
\label{wave2}
\Box \phi=U'(\phi)\,,
\ee
where $U$ denotes the potential of the scalar field and a prime denotes differentiation with respect to the argument. The assumptions regarding symmetries and asymptotics are identical to those above. However, if one wants to have $\phi \to \phi_0$ at infinity in this case, then the potential has to satisfy $U'(\phi_0)=0$, and this will be an extra requirement. Note that this requirement implies that the potential should either have an extremum, or be a runaway potential with a flat end. Now, $\phi=\phi_0$ is obviously a solution and the idea is to show that it is unique. To this end, suppose that $\phi \neq \phi_0$ and $U'\neq 0$.\footnote{One could have a situation where there are more than one values of the scalar for which $U'$ vanishes. Since $\phi_0$ is not assumed to have any specific value it can collectively denote all constant $\phi$ solutions.} Multiplying by $U'$ and integrating over ${\cal V}$ as above yields
\begin{equation}
\int_{\cal V} d^4 x\sqrt{-g}\, U'(\phi)\Box \phi= \int_{\cal V} 
d^4 x\sqrt{-g}\, U'^2(\phi).
\end{equation}
Integrating by part one gets
\begin{eqnarray}
\label{int}
\int_{\cal V} d^4 x\sqrt{-g}\, \big[U''(\phi)\nabla^\mu \phi\nabla_\mu \phi + U'^2(\phi)\big]=\nonumber\\=\int_{\partial{\cal V}} d^3 x\sqrt{|h|}\, U'(\phi)n^\mu\nabla_\mu\phi \,,
\end{eqnarray}

The boundary contribution on the right hand side vanishes for precisely the same reasons as in the case where $U=0$. On the other hand, the integrand of the left hand side is manifestly positive as long as $U''\geq 0$. So, as long as $U''\geq 0$ everywhere in spacetime, the only way for the integral to vanish is if $\phi=\phi_0$ (in which case $U'(\phi_0)=0$ as well). The $U'' \geq 0$ condition can be interpreted as a local stability condition for the scalar field:  consider a small enough patch of spacetime so that curvature effects can be neglected and linearise eq.~(\ref{wave2}) in a flat background and around an arbitrary solution. When $U''<0$ the perturbations will exhibit a tachyonic instability.

\section{Scalar-tensor theories}
\label{ST}
\subsection{No-hair theorem}
\label{sttheorem}

So far we have only considered a scalar field on a black hole background. However, it is straightforward to show that the results of the previous section are directly applicable to scalar-tensor theories of gravity. The latter are described by the action
\be
\label{staction}
S_{\rm st}=\int d^4x \sqrt{-\hat{g}} \Big(\varphi 
\hat{R}-\frac{\omega(\varphi)}{\varphi} \hat{\nabla}^\mu \varphi 
\hat{\nabla}_\mu \varphi -V(\varphi) 
+L_m(\hat{g}_{\mu\nu},\psi)\Big)\,,
\ee
which is written is the {\em Jordan frame}, so $\hat{g}_{\mu\nu}$ is the Jordan frame metric, $\hat{R}$ the associated Ricci scalar, and $\hat{\nabla}_\mu$ the covariant derivative defined with this metric. $\varphi$ is a scalar field, $L_m$ is the matter Lagrangian and $\psi$ collectively denotes the matter fields. $\hat{g}_{\mu\nu}$ is understood to couple minimally to the matter fields and $\varphi$ does not enter the matter action. These last two requirements define the Jordan frame. One might be tempted to think that this action might be generalised further by turning the coefficient of $\hat{R}$ into an arbitrary function of $\varphi$, but this just amounts to a scalar field redefinition \cite{Flanagan:2003iw, Sotiriou:2007zu}.

The conformal transformation $g_{\mu\nu}=\varphi 
\, \hat{g}_{\mu\nu}$ followed by the scalar field redefinition
\be
\label{redef}
d\phi=\sqrt{\frac{2\omega(\varphi)+3}{16\pi}} \, d\varphi\,,
\ee
brings the action~(\ref{staction}) into the form
\be
\label{stactionein}
S_{\rm st}=\int d^4x \sqrt{-g} \Big(\frac{R}{16\pi}-\frac{1}{2} \nabla^\mu \phi \nabla_\mu \phi -U(\phi)+L_m(\hat{g}_{\mu\nu},\psi)\Big)\,,
\ee
where  $U(\phi)=V(\varphi)/\varphi^2$. This representation of scalar-tensor theory is called the {\em Einstein frame} and $g_{\mu\nu}$ and $\phi$ are the Einstein frame metric and scalar field respectively. In this representation $\phi$ couples minimally to gravity and has a canonical kinetic term, so the theory resembles general relativity. The trade-off is that the scalar $\phi$ now couples to matter.

In vacuo, variations with respect to $g^{\mu\nu}$ and $\phi$ yield the Einstein frame field equations
\begin{eqnarray}
\label{field1}
R_{\mu\nu}-\frac{1}{2} R g_{\mu\nu}&=&8\pi\,T_{\mu\nu}^\phi\,,\\
\label{field2}
\Box \phi= U'(\phi)\,,
\end{eqnarray}
where
\be
T_{\mu\nu}^\phi=\nabla_\mu \phi \nabla_\nu \phi 
-\frac{1}{2}g_{\mu\nu}\nabla_\lambda \phi \nabla^\lambda \phi 
-U(\phi)g_{\mu\nu}\,.
\ee
Eq.~(\ref{field2}) is no different than eq.~(\ref{wave2}) so the analysis of the previous section that proves that $\phi=\phi_0$ is the unique solution still applies, with one subtlety. The metric is not assumed to satisfy vacuum Einstein equations but instead it satisfies eq.~(\ref{field1}). An element of the proof was that stationarity implied axisymmetry by virtue of Hawking's theorem \cite{Hawking:1971vc}. This theorem assumes the Weak Energy Condition, $T_{\mu\nu} \chi^\mu \chi^\nu \geq 0$ for any timelike vector $\chi^\mu$. Applied on  $T_{\mu\nu}^\phi$ this condition reads 
\be
\label{wec}
(\chi^\mu \nabla_\mu \phi)^2 
-\frac{1}{2}\chi^2\nabla_\lambda \phi \nabla^\lambda \phi 
-U(\phi)\chi^2\geq 0\,,
\ee
where $\chi^2=g_{\mu\nu}\chi^\mu\chi^\nu<0$, as $\chi^\mu$ is a timelike vector. On the other hand $\nabla_\lambda \phi \nabla^\lambda \phi>0$. So, the first two terms in the left hand side of eq.~(\ref{wec}) are manifestly positive and the Weak Energy Condition holds provided that $U$ does not became overly negative. As long as this requirement and the assumptions laid out in the previous section are satisfied, then $\phi=\phi_0$ and $T_{\mu\nu}^\phi= 
-U(\phi_0)g_{\mu\nu}$. Since the solution has to be asymptotically flat by assumption, $U(\phi_0)=0$ and the metric satisfies vacuum Einstein's equations. Hence, the solutions will be identical to those of general relativity.

Before closing this section, it is worth mentioning that, if one is willing to restrict one's attention to static, spherically symmetric solution, then there exist different proofs of no-hair theorems. In some cases, these proofs are applicable to more general theories than standard scalar-tensor theories ({\em e.g.}~complex scalars or scalars coupled to gauge fields). See Ref.~\cite{Bekenstein:1996pn} and references therein.

\subsection{Significance and limitations of no-hair theorems}
\label{lims}

The no-hair theorems outlined above  suggest that quiescent (stationary), isolated (asymptotically flat) black holes in scalar tensor theory will not differ from black holes in general relativity.  However, any theorem is as good as its assumptions, so a critical discussion about these assumptions is due.

\subsubsection{Quiescent versus perturbed black holes:} 

The assumption of stationarity is a quite reasonable one when one is interested in the final state of gravitational collapse. However, an astrophysical black holes will cease to be stationary once it is perturbed. Scalar-tensor theories have an extra scalar degree of freedom with respect to general relativity, so there is an extra mode that can be excited. Hence, perturbation theory around the same solution in general relativity and in scalar tensor theory will differ. This is rather important, as it opens a window for detecting deviations from general relativity by studying black holes despite the existence of no-hair theorems \cite{Barausse:2008xv}. In fact, the existence of a scalar mode can potentially lead to smoking gun effects in astrophysical systems (such as ``floating orbits'' for example \cite{pressteu,Cardoso:2011xi}). Several contributions to this volume will be devoted to such perturbative effects, so we will refrain from saying more here. The significance of the no-hair theorems in this context is that they pinpoint the background one needs to consider in black hole perturbation theory.

\subsubsection{Asymptotics:}

Asymptotic flatness is a crucial assumption for no-hair theorems. Hairy black holes with anti-de Sitter (AdS) asymptotics do exist in scalar tensor theory, see for example Ref.~\cite{Torii:2001pg}.  For de Sitter asymptotics one could generalise the theorems presented above quite trivially. The way the asymptotics enter the analysis is the assumption that part of the boundary $\partial {\cal V}$ of the region ${\cal V}$ under consideration is a timelike 3-surface at infinity. Asymptotic flatness is then used to argue that the contribution to the integral from this part of the boundary vanishes in eq. (\ref{int}). In the case of de Sitter asymptotics one could replace the timelike 3-surface at infinity with a part of the de Sitter horizon. The contribution from this part would also vanish, as the gradient of the scalar would be normal to the generator of the de Sitter horizon by symmetry assumptions (same as it is for the black hole horizon). 

More realistically though, one might be interested in what happens when the black holes is ``embedded'' in an evolving universe, {\em i.e.}~one actually  has Friedmann-Lema\^itre--Robertson--Walker asymptotics. Generically, this would introduce an evolution to the scalar field and one could expect hair to form as a result of this evolution. However, it is also natural to think that such effects would be naturally suppressed by the dimensionless ratio $t_{\rm sys}/t_{\rm cosm}$, where $t_{\rm sys}$ is the characteristic timescale of the system in consideration and $t_{\rm cosm}$ is the time scale of cosmic evolution. These scales are very well separated, with the exception of primordial black holes in the early universe \cite{Jacobson:1999vr}. Hence, one would expect that the induced hair is quite hard to detect. This expectation is indeed supported by more robust calculations \cite{Horbatsch:2011ye}.

Asymptotics also seem to impose rather severe restriction to the form of the potential of the scalar one can consider as, in order for flat space to be admissible when the scalar takes its asymptotic value $\phi_0$, one needs $U(\phi_0)=0$. However, when viewed in the context of separation of scales discussed above, this restriction might not be that severe. $U(\phi_0)$ effectively behaves like a cosmological constant, so as long as it is small enough one should be able to safely neglect it for the same reasons one is neglecting the cosmic evolution. After all, asymptotic flatness is in this context just a mathematical idealisation that reflects the physical assumption that the system is (relatively) isolated.

\subsubsection{Symmetries and the scalar field:} When one imposes that a spacetime has certain symmetries it is tempting to assume that not only the metric, but also the rest of the fields inherit these symmetries. This is certainly the assumption made in the no-hair theorems, as the scalar is taken to have vanishing Lie derivatives with respect to the Killing vectors of the metric. However, it is not inconceivable that  solutions where the scalar has less symmetry than the spacetime are admissible. The field equations do not actually imply that the scalar has to respect the symmetries of the metric. For example, take eq.~(\ref{field1}): the left hand side will have vanishing Lie derivative with respect to any Killing vector of the metric, which implies that the same should be true for $T_{\mu\nu}^\phi$. But this is a far weaker restriction than asking for $\phi$ itself to have vanishing Lie derivative. It has recently been claimed in Ref.~\cite{Graham:2014ina} that  stationary black holes with a time dependent scalar field do not actually exist so long as the scalar field is real. On the contrary, for complex scalar fields (which we have not considered above) hairy solutions in which the scalar field is time-dependent but the metric is stationary do exist \cite{Herdeiro:2014goa}. Whether or not such solution can arise as the outcome of gravitational collapse is still unclear. This question is particularly relevant in the present context, as stationarity is the mathematical assumption that reflects our desire to determine the final state of collapse.

\subsubsection{Surrounding matter:} Black holes are vacuum solutions. However, realistic astrophysical black holes have matter in their vicinity in the form of an accretion disk, companion stars or a galactic environment. It is thus pertinent to understand how this surrounding matter affects the configuration of the scalar field. When matter is taken into account eq.~(\ref{field2}) acquires an extra contribution and takes the form
\be
\Box \phi= U'(\phi)-A'(\phi) A^3(\phi) T\,,
\ee
where $A(\phi)=\varphi$ is the conformal factor that relates the Jordan and the Einstein frame metrics and $T$ is the trace of the stress-energy tensor of matter (as defined in the Jordan frame). For a given $\omega(\varphi)$, $A(\phi)$ can be obtained by integrating eq.~(\ref{redef}), so it  is essentially a free function that defines the theory. It is easy to see that, so long as $T\neq 0$, $\phi=\phi_0=$constant solutions are only admissible if $U'(\phi_0)=A'(\phi_0)=0$. ($A(\phi_0)$ cannot vanish, as it is the conformal factor of the metric.) This means that $U$ and $A$ should have stationary points for the same value of $\phi$, which is a rather stringent restriction. Even for a vanishing potential $U$,  $A$ does not  have to have a stationary point at all in general. Remarkably, for any theory that does not satisfy this requirement, even a tiny amount of matter would  force the scalar field to acquire a non-trivial configuration and induce hair to the black hole. Of course the important, and so far largely unexplored, question is if realistic  matter configurations around astrophysical black holes induce detectable hair.

The theories for which $A$ does have a stationary point are special and have been studied extensively, mostly in the absence of a a potential $U$. It is clear that when $A'(\phi_0)=0$ the corresponding theory admits all of the spacetimes of general relativity as solutions even in the presence of matter. Hence, one might expect that such theories can evade weak field experiments and this is indeed the case. Interestingly though, they can still be significantly constrained by strong gravity tests due to a phenomenon called spontaneous scalarization \cite{Damour:1993hw,Damour:1996ke}. The basis of this effect is the fact that even though the solution of general relativity with constant scalar are in principle always admissible, they are not unique. For stars with relatively low central densities, such as the Sun, they seem to be indeed energetically preferable. However, as the compactness of the star increases a transition occurs and above a certain threshold a ``scalarized" configuration is the final result of gravitational collapse. 

At the perturbative level, spontaneous scalarization tends to manifest as a tachyonic instability of the unscalarized solution. The coupling between the scalar and the matter render the scalar perturbation massive and the induced effective mass is negative. 
It has recently been shown in Refs.~\cite{Cardoso:2013fwa,Cardoso:2013opa} that spontaneous scalarization can occur in black holes as well when the matter distribution that surrounds the black hole induces a negative effective mass. On the other hand, when the effective mass turns out to be positive and the black hole is rapidly rotating other interesting phenomena can occur, such us superradiance driven instabilities and large amplification \cite{Cardoso:2013fwa,Cardoso:2013opa}. The relevance of these effects for realistic matter configuration is still unclear and deserves further investigation.

\section{Generalized scalar-tensor theories}
\label{othertheories}

The most obvious way to evade the theorems presented in Sections \ref{theorems} and \ref{sttheorem} is to consider more general actions that contain a scalar field non-minimally coupled to gravity. The theorems do not apply to theories were the scalar does not satisfy eq.~(\ref{wave2}) (or cannot be made to do so by means of suitable field redefinitions). For example, it has been known for quite some time \cite{Kanti:1995vq} that static, spherically symmetric black holes do have hair in the following theory:
 \be
\label{gbaction}
S_{GB}=\frac{1}{2}\int d^4x \sqrt{-g} \left( R - \frac{1}{2}\partial_\mu \phi \partial^\mu \phi +\alpha e^{\phi} \mathcal{G} \right)\,,
\ee
where ${\cal G}\equiv R^{\mu\nu\lambda\kappa}R_{\mu\nu\lambda\kappa}-4 R^{\mu\nu}R_{\mu\nu}+R^2$ is the Gauss--Bonnet invariant. It is worth stressing that one cannot ``undo'' the coupling between the scalar and ${\cal G}$  by means of a conformal transformation. One could attempt to redefine $\phi$ in order to change the coupling's dependence on $\phi$ but any such redefinition would make the kinetic term non-canonical. So, it should be clear that the scalar field in this theory cannot be made to satisfy eq.~(\ref{wave2}) by means of a field redefinition.

More recently it has been shown that black holes have hair in a similar theory with a linear coupling between $\phi$ and ${\cal G}$ \cite{Sotiriou:2013qea,Sotiriou:2014pfa}
 \be
\label{gblaction}
S_{GBL}=\frac{1}{2}\int d^4x \sqrt{-g} \left( R - \frac{1}{2}\partial_\mu \phi \partial^\mu \phi +\alpha \phi \mathcal{G} \right)\,.
\ee
In this theory it is straightforward to see that black holes {\em have to} have hair by simply inspecting the equation for the scalar
\be
\label{gbeq}
\Box \phi+\alpha {\cal G}=0\,.
\ee
Since ${\cal G}$ contains the Kretschman scalar, $R^{\mu\nu\kappa\lambda}R_{\mu\nu\kappa\lambda}$, that vanishes only in flat spacetime, it is clear the $\phi$ cannot be constant in any other spacetime \cite{Sotiriou:2013qea}. Actually, the same argument can be made for any theory with a coupling of the type $f(\phi) {\cal G}$ for which $f'$ never vanishes.

The hairy black hole solutions of Refs.~\cite{Kanti:1995vq} and \cite{Sotiriou:2014pfa} have some remarkable features. Even though they do have a non-trivial scalar configuration their scalar charge is not an independent parameter, so one still has just a one-parameter family of solutions (the parameter can be related to the ADM mass). For the case of action (\ref{gblaction}) the reason can already be seen if one tries to solve eq.~(\ref{gbeq}) on a Schwarzschild background perturbatively in orders of the coupling $\alpha$, or more correctly the dimensionless quantity $\alpha/m^2$ where m is the mass of the black hole and $\alpha/m^2\ll 1$ \cite{Sotiriou:2013qea} (see also Refs.~\cite{Mignemi:1992pm,Yunes:2011we}). To first order in $\alpha/m^2$ the solution is 
\be
\phi'=\alpha{\frac {16{m}^{2}-C {r}^{3}}{{r}^{4} \left( r-2m \right)}}    
\ee
where $r$ is the areal radius coordinate and $C$ an integration constant. $\phi'$ diverges on the black hole horizon $r=2m$ and the only way to avoid this is to impose $C=2/m$. This choice yields
\be
\phi'=-\frac{2 \alpha}{m}{\frac {(r^2+2m r +4m^2)}{{r}^{4} }} \,.
\ee
Hence, regularity of the scalar across the black hole horizon fixes the value of $C$ and relates the scalar charge with the black hole mass. Non-trivial scalar profiles whose scalar charge is not an independent parameter are sometimes referred to as ``hair of the second kind''.

The other remarkable feature of these hairy solutions, which is not captured by the perturbative treatment in the coupling, is that they harbour finite area instead of central singularities. The size of the finite area singularity is determined not only by the mass but also the value  of the coupling $\alpha$, and this implies that such black holes have a minimum size.  It is not surprising that this feature of the geometry is not seen in the perturbative treatment used above. If one calculates when the perturbative treatment breaks down by comparing the size of the leading order term in the expansion with the size of the next-to-leading-order correction, one finds that they become comparable at precisely the radius were the finite area singularity lies in the non-perturbative solutions \cite{Sotiriou:2014pfa}. Note that in some cases it might not be wise to take the non-perturbative solution seriously. For instance, suppose one is considering $\alpha \phi {\cal G}$ as the first in a series of corrections to the action where $\alpha$ is the parameter controlling the order of the expansion. Then the solution can only be trusted to order $\alpha$ (and only if it is smooth in the $\alpha \to 0$ limit) as ${\cal O}(\alpha^2)$ correction have been neglected at the level of the action (see Ref.~\cite{Sotiriou:2014pfa} for a more detailed discussion).

It should be clear by now that one cannot hope to have a no-hair theorem that applies to any generalised scalar-tensor theory. However, it is worth exploring whether such a theorem can be proven for at least a subset of the possible generalisations.


\subsection{No-hair theorem in shift-symmetric Horndeski theory}

Horndeski in 1974 has pinned down the most general action that leads to second order equations for a metric and for a scalar field in 4 dimensions \cite{Horndeski:1974wa}. This result had passed largely unnoticed until fairly recently, when Horndeski's theory got rediscovered \cite{Deffayet:2011gz} as a covariant generalisation of Galileon's \cite{Nicolis:2008in} --- flat spacetime scalar theories where the scalar enjoys the galilean-like symmetry $\phi \to \phi + c_\mu x^\mu +c$, where $c_\mu$ is a constant one-form, $c$ a constant and $x^\mu$ is the coordinate vector. Galilean symmetry is absent in curved space and only a subclass on Horndeski's theory actually enjoys it in the flat spacetime limit. We will not delve any deeper in the general properties of Horndeski theory or Galileons here. For a recent review please see Ref.~\cite{Deffayet:2013lga}. 

If one is willing to restricts oneself to scalar fields that enjoy shift symmetry, {i.e.}~invariance under $\phi \to \phi +c$ then the restricted Horndeski action takes the form \cite{Sotiriou:2014pfa}
\begin{eqnarray}
\label{ssgg}
{\cal L}&=& {\cal  L}_2+{\cal  L}_3+{\cal  L}_4+{\cal L}_5 ,
\\
\label{ssggl2}
{\cal L}_2 &=& K(X) ,\nn
\\ 
{\cal L}_3 &=&-G_3(X) \Box \phi ,\nn
\\
{\cal L}_4 &=& G_4(X) R + G_{4X} \left[ (\Box \phi)^2 -(\nabla_\mu\nabla_\nu\phi)^2 \right] ,\nn
\\
\label{ssggl5}
{\cal L}_5 &=& G_5(X) G_{\mu\nu}\nabla^\mu \nabla^\nu \phi - \frac{G_{5X}}{6}  \big[ (\Box \phi)^3
- 3\Box \phi(\nabla_\mu\nabla_\nu\phi)^2 + 2(\nabla_\mu\nabla_\nu\phi)^3 \big] ,\nn
\end{eqnarray}
where $K, G_3, G_4, G_5$ are arbitrary functions of $X\equiv- \partial^\mu \phi \partial_\mu \phi/2$. Additonally, $f_X\equiv\partial f(X)/\partial X$, $G_{\mu\nu}$ is the Einstein tensor, $(\nabla_\mu\nabla_\nu\phi)^2\equiv \nabla_\mu\nabla_\nu\phi \nabla^\nu\nabla^\mu\phi$ and $(\nabla_\mu\nabla_\nu\phi)^3=\nabla_\mu\nabla_\nu\phi \nabla^\nu\nabla^\rho\phi \nabla_\rho\nabla^\mu\phi$. Shift symmetry implies that the equation of motion for the scalar take the form of a current-conservation equations
\be
\label{eqphi}
\nabla_\mu J^\mu=0\,,
\ee
where $J^\mu$ is the Noether current associated with shift symmetry. 
 In Ref.~\cite{Hui:2012qt} a no-hair theorem for the shift-symmetric Hornedeski action has been presented. The theorem applies to asymptotically flat black holes and the symmetry assumptions are stricter that those of the no-hair theorems in standard scalar-tensor gravity. Instead of just stationarity, staticity and spherical symmetry are assumed. Finally, there is also an assumption about the precise form of the current as a functional of $\phi$ which, as we will see below, will restrict the range of validity of the theorem.
 
Under the assumptions of staticity and spherical symmetry one is allowed to make the ansatz 
\be
\label{mansatz}
ds^2=-f(\rho)dt^2+f(\rho)^{-1}d\rho^2+r^2(\rho)d\Omega^2
\ee
for the metric. Assuming that $\phi$ respects the same symmetries one has $\phi=\phi(\rho)$. The proof can then be split into the following steps:
 \begin{itemize}
 \item $J^\rho$ should be the only non-vanishing component of the current. A non-vanishing angular component is excluded by spherical symmetry and $J^t$ has to vanish else it would imply the existence of a preferred time direction.
 \item $J^\rho$ has to vanish on the black hole horizon else $J^\mu J_\mu= (J^\rho)^2/f$ would diverge there. Note that $f$ is the norm of the Killing vector associated with symmetry under time translations, so it vanishes on the horizon.  
 \item $J^\rho$ has to vanish everywhere if it vanishes on the horizon.  Integrating eq.~(\ref{eqphi}) yields $r^2(\rho) J^\rho=\;$constant and $r$ is the areal radius of constant-$\rho$ surfaces, so it is finite on the horizon.
 \item $J^\rho=0$ implies that $\phi$ is constant. 
 \end{itemize}
 It is clear that this last step is non-trivial and depends on how $\phi$ and its derivatives enter the current. In Ref.~\cite{Hui:2012qt} it has been assumed that the 
 \be
 \label{Jrho}
J^\rho=f\,\partial_\rho\phi F(\partial_\rho\phi\, ; g, \partial_\rho g, \partial_\rho\partial_\rho g)\,.
\ee
where $F$ tends to a non-zero constant asymptotically. The justification for the ansatz and the asymptotic behaviour of $F$ is that the leading order term in the weak field limit, which should be applicable asymptotically, should be coming from the standard canonical kinetic term. It is then argued that, since $f=1$, $F\neq 0$ and $\partial_\rho\phi=0$ asymptotically and $f$ and $F$ are smooth functions, as ones moves infinitesimally closer to the black hole $J^\rho$ cannot remain zero unless $\partial_\rho\phi$ remains zero as well. 

The hairy black hole solutions to the theory (\ref{gblaction}) discussed above seem to already constitute counter examples to this no-hair theorem \cite{Sotiriou:2013qea}. Action (\ref{ssgg}) is supposed to be the most general action for a shift symmetric scalar that leads to second order field equations for both the scalar and the metric. Action (\ref{gblaction}) is shift-symmetric up to a boundary term, as ${\cal G}$ is a total divergence, and it certainly leads to 2nd order equations. So, it has to be a subcase of (\ref{ssgg}). Though not obvious, this is indeed the case and one needs to make the choice $K=X$, $G_3=0$, $G_4=M_p^2/2$, and $G_5=-4\alpha \ln|X|$ in order to recover (\ref{gblaction}) from (\ref{ssgg}) \cite{Kobayashi:2011nu}. 

The way this specific choice evades the no-hair theorem is by failing to satisfy the assumption made for the form of the current (see Ref.~\cite{Sotiriou:2013qea} for a more detailed discussion on the form of the current). Reading off the current from eq.~(\ref{gbeq}) yields
\be
J^{\mu}=\nabla^\mu \phi- \nabla^\mu {\cal G}_\mu\,,
\ee
where ${\cal G}_\mu$ can be thought of as implicitly defined by $\nabla^\mu {\cal G}_\mu={\cal G}$. The second term depends on the metric only and so $J^\rho$ cannot be put in the form of the ansatz (\ref{Jrho}) with $F$ being continuous as $\partial_\rho\phi\to 0$. 

As it has been argued in Ref.~\cite{Sotiriou:2013qea}, $\phi {\cal G}$ is the only term contained in the action (\ref{ssgg}) that can evade the theorem of Ref.~\cite{Hui:2012qt} and still admit flat spacetime with constant $\phi$ as a solution. This is because, the only way that the current can be finite asymptotically, where the spacetime is flat and $\partial_\rho\phi\to 0$, and not satisfy the assumptions of Ref.~\cite{Hui:2012qt} is if it has a piece that is independent of $\phi$. This in turn implies that there is a term in the action that is linear in $\phi$, {\em i.e.}~it has the form $\phi A[g]$. The function of the metric $A$ should be a total divergence else the action would fail to be shift-symmetric. Additionally, variation of this term with respect to the metric and the scalar should lead to contributions to the field equations that are second order in derivatives. Combining all of these requirements one gets that $A[g]={\cal G}$.

It should be clear that, as long as the term $\phi {\cal G}$ is not excluded by fiat from the action (\ref{ssgg}), black hole solutions will have hair irrespectively of what other terms might be present. Recall that ${\cal G}$ only vanishes in flat space and it will be effectively sourcing $\phi$ in any theory that contains the term $\phi {\cal G}$. One could exclude $\phi {\cal G}$ from the action by imposing some extra internal symmetry \cite{Sotiriou:2013qea}.  The most obvious would be $\phi\to-\phi$ and the corresponding theory would be
\be
\label{lagrsym}
{\cal L}=K(X) + G_{4}(X)R+G_{4X}\left[
\left(\Box\phi\right)^2-\left(\nabla_\mu\nabla_\nu\phi\right)^2\right]\,.
\ee

\subsection{Limitation and extensions of the no-hair theorem}

The most obvious limitation of the no-hair theorem for shift-symmetric Horndeski theory is the loophole associated with the term $\phi {\cal G}$ that has already been discussed. In absence of this term, either by fiat or by assuming extra symmetries that exclude it, the theorem will apply and black holes will no differ from those of GR. However, all of the limitations listed in Section \ref{lims} for standard scalar tensor theory apply here as well. We will not repeat the discussion but it is worth emphasising that the assumptions regarding asymptotics and symmetries are even more subtle here. Because of shift symmetry the scalar appears in the equations only through its gradient, so there is a priori no reason to assume that $\phi$ itself respects the symmetries of the metric. It would be more than enough to just assume that the gradient of $\phi$ has vanishing Lie derivatives with respect to the Killing vectors \cite{Sotiriou:2013qea}. Indeed, hairy static black hole solutions with a time-dependent scalar do exist \cite{Babichev:2013cya}. Additionally, since theories described by the action (\ref{ssgg}) can lead to rather non-trivial cosmologies, it is worth studying in detail how time-dependent asymptotic conditions for the scalar would affect its configuration. 

One extra limitation of this no-hair theorem with respect to those of standard scalar-tensor theory is that it resorts to staticity and spherical symmetry, as opposed to just stationarity. Because of these extra symmetry assumptions, it does not cover rotating black holes. However, it is rather straightforward to extend its validity to slow rotation \cite{Sotiriou:2013qea}. Let us assume that starting from a static, spherically symmetric solution one can generate a rotating one perturbatively order by order in the rotation, {i.e.~}a single book keeping parameter
$\epsilon$ is enough to keep track of the corrections. Changing the direction of rotation would imply $\epsilon\to -\epsilon$. One can argue one physical grounds that the solution should be invariant under reversal of the direction of rotation together with reversal of either the time coordinate or the azimuthal angle. The most general slowly rotating metric has been given on the basis of this argument in Ref.~\cite{hartle67}. One could apply this argument to the scalar $\phi$. The slowly rotating solution will be stationary and axisymmetric. In the coordinate system used to write the metric ansatz (\ref{mansatz}) one then has $\phi=\phi(\rho,\theta)$, as $\phi$ is assumed to respect the symmetries of the metric. Corrections linear in the rotation would change sign if the direction of rotation is reversed irrespectively of whether one reverses time or azimuthal angle. Therefore, they have to be absent and $\phi$ can receive no correction at linear order in rotation. As a result, given that $\phi=$constant  in the spherical case it will have to remain so for slowly rotating solutions, and the corresponding metric will solve Einstein's equations. 

\subsection{Theories with a preferred foliation}

There exist generalised scalar-tensor theories which have second order field equation and yet they are not members of the Horndeski action discussed above. This might seem as an apparent contradiction, as Horndeski's action is by construction the most general one that leads to second order equations for both the metric and the scalar. The subtlety lies on the following fact: Horndeski's requirement is that the equations are second order in every foliation whereas one could actually relax this requirement and request instead that there is at least one foliation in which the equations are second order. Theories of this second type would clearly have a preferred foliation. In their covariant formulation they would have higher than second order derivatives.

Consider in particular the case of a scalar field $\phi$ that appears in the action only through the combination $u_\mu= N \partial_\mu \phi$, where $N= (g^{\mu\nu}\partial_\nu \phi \partial_\mu \phi)^{-1/2}$. $u^\mu$ is then by definition a unit, timelike vector and the scalar will have to have a timelike gradient in every solution. As a consequence, constant $\phi$ surfaces will define a foliation and $N$ will be the lapse of this foliation. If one constructs a theory which is second order in derivatives of  $u_\mu$ then it will be inevitably third order in derivatives of $\phi$ (which is the fundamental field here). However, if one chooses $\phi$ as a time coordinate $T$, which is a reasonable choice as $\phi$ has a timelike gradient, then $u_\mu=N \delta^T_\mu$. In this preferred foliation $u_{\mu}$ has ceased to carry any derivatives in its definition, so the equations will be second order if the theory was constructed to have two derivative of $u_\mu$.

The structure just described is present in the low-energy limit of Ho\v rava gravity \cite{Horava:2009uw,Blas:2009qj}, see Ref.~\cite{Jacobson:2010mx} for details. The scalar field $\phi$ can be seen as dynamically defining the preferred foliation in each solution. It should be clear that black hole solutions in this theory, and in general in theories where the scalar defines a preferred foliation, will always have hair, as the scalar can never been in a trivial ($\phi=$constant) configuration. Static, spherically symmetric solution have been studied in Ref.~\cite{Barausse:2011pu} and remarkably the scalar charge is not an independent parameter if the scalar, or actually $u_\mu$, is to be regular everywhere apart from the central singularity. 

The structure of black holes in theories with a preferred foliation and the associated phenomenology is expected to be very different than that of general relativity or the rest of the theories considered above. Hence, we will not discuss them any further here and we refer the reader to a recent review and references therein instead \cite{Barausse:2013nwa}. The most remarkable feature of such black holes is that they can harbor a {\em universal horizon}: a surface that traps any signal, irrespectively of how fast it propagates \cite{Barausse:2011pu,Blas:2011ni,Barausse:2012ny,Barausse:2012qh,Berglund:2012bu,Sotiriou:2014gna,Bhattacharyya:2014kta}. This is a particularly relevant concept in theories with a preferred foliation, as in such theories one can have superluminal and even instantaneous propagation, so the existence of universal horizons implies that the notion of a black hole actually survives.

\section{Conclusions}
\label{concl}

We have considered the effect a scalar field can have in the structure of black holes. No-hair theorems indicate that stationary, asymptotically flat black holes in standard scalar tensor theory will have to have constant scalar configurations and will therefore be solution of general relativity. The same statement can be made for static, spherically symmetric, asymptotically flat black holes and their slowly rotating counterpart in the most general shift-symmetric scalar-tensor theory that leads to second order field equations, provided that a linear coupling between the scalar and the Gauss--Bonnet invariant is excluded by fiat. In all other theories, scalar fields will generically have a non-trivial configuration in a  black hole spacetime and will lead to a different black hole structure than that of general relativity.

No-hair theorems are elegant and powerful, in the sense that they make very few assumptions and yet they are applicable to a rather wide class of theories. However, one has to interpret them with caution. Stationarity and asymptotic flatness, for instance, might seem as very minimal and rather realistic assumptions but they are at best approximate in nature. As has been discussed above in detail, astrophysical black holes might be surrounded by matter in the form of a companion star, an accretion disk, or a galaxy, which would inevitably introduce scalar hair. Time-dependent cosmological asymptotics can have the same effect. Moreover, in certain case one can find hairy solution where the metric satisfies the assumptions of the no-hair theorems but the scalar does not. Finally, black hole perturbation theory is certainly different in (generalised) scalar-tensor theories than in general relativity.  When all of these facts are taken into consideration it becomes obvious that realistic, astrophysical black holes in (generalised) scalar-tensor theories will certainly differ from their general relativity counterparts. The important open question, which is currently under scrutiny and will be discussed in many contribution to this special issue, is how detectable are these deviations.

\ack

I am grateful to Paolo Pani and Helvi Witek for a critical reading of this manuscript and useful suggestions.
The research leading to these results has received funding from the European Research Council under the European Union's Seventh Framework Programme (FP7/2007-2013) / ERC Grant Agreement n.~306425 ``Challenging General Relativity''. Research at Perimeter Institute is supported by the Government of Canada through Industry Canada and by the Province of Ontario through the Ministry of Economic Development \& Innovation.


\section*{References}

\begin{thebibliography}{99}

\bibitem{Hawking:1971vc} 
  S.~W.~Hawking,
  Commun.\ Math.\ Phys.\  {\bf 25}, 152 (1972).
  

\bibitem{israel1} 
  W.~Israel,
  Phys.\ Rev.\  {\bf 164}, 1776 (1967).

\bibitem{israel2} 
  W.~Israel,
  Commun.\ Math.\ Phys.\  {\bf 8}, 245 (1968).

\bibitem{Carter:1971zc} 
  B.~Carter,
  Phys.\ Rev.\ Lett.\  {\bf 26}, 331 (1971).

\bibitem{Wald:1971iw} 
  R.~M.~Wald,
  Phys.\ Rev.\ Lett.\  {\bf 26}, 1653 (1971).
  
  \bibitem{wheeler}
R.~Ruffini and J.~A.~Wheeler, Physics Today 24, 30 (1971).  

\bibitem{Volkov:1998cc} 
  M.~S.~Volkov and D.~V.~Gal'tsov,
  Phys.\ Rept.\  {\bf 319}, 1 (1999)
  [hep-th/9810070].
  
\bibitem{Sotiriou:2015lxa} 
  T.~P.~Sotiriou,
  Lect.\ Notes Phys.\  {\bf 892}, 3 (2015)
  [arXiv:1404.2955 [gr-qc]].
  
\bibitem{Hawking:1972qk} 
  S.~W.~Hawking,
  Commun.\ Math.\ Phys.\  {\bf 25}, 167 (1972).
  
\bibitem{Sotiriou:2011dz} 
  T.~P.~Sotiriou and V.~Faraoni,
  Phys.\ Rev.\ Lett.\  {\bf 108}, 081103 (2012)
  [arXiv:1109.6324 [gr-qc]].
  
\bibitem{Flanagan:2003iw}
  E.E.~Flanagan,
Class.\ Quant.\ Grav.\ {\bf 21}, 417 (2003)  [gr-qc/0309015].
  
\bibitem{Sotiriou:2007zu}
  T.P.~Sotiriou, V.~Faraoni, and S.~Liberati,
 Int.\ J.\ Mod.\ Phys.\ D {\bf 17}, 399 (2008)
  [arXiv:0707.2748 [gr-qc]].

\bibitem{Bekenstein:1996pn} 
  J.~D.~Bekenstein,
  In *Moscow 1996, 2nd International A.D. Sakharov Conference on physics* 216-219
  [gr-qc/9605059].

\bibitem{Barausse:2008xv} 
  E.~Barausse and T.~P.~Sotiriou,
  Phys.\ Rev.\ Lett.\  {\bf 101}, 099001 (2008)
  [arXiv:0803.3433 [gr-qc]].
  
  \bibitem{pressteu}   W.~H.~Press and S.~A.~Teukolsky, Nature {\bf 238}, 211 (1972).
  
\bibitem{Cardoso:2011xi} 
  V.~Cardoso, S.~Chakrabarti, P.~Pani, E.~Berti and L.~Gualtieri,
  Phys.\ Rev.\ Lett.\  {\bf 107}, 241101 (2011)
  [arXiv:1109.6021 [gr-qc]].
 
\bibitem{Torii:2001pg} 
  T.~Torii, K.~Maeda and M.~Narita,
  Phys.\ Rev.\ D {\bf 64}, 044007 (2001).

\bibitem{Jacobson:1999vr} 
  T.~Jacobson,
  Phys.\ Rev.\ Lett.\  {\bf 83}, 2699 (1999)
  [astro-ph/9905303].

\bibitem{Horbatsch:2011ye} 
  M.~W.~Horbatsch and C.~P.~Burgess,
  JCAP {\bf 1205}, 010 (2012)
  [arXiv:1111.4009 [gr-qc]].
  
\bibitem{Graham:2014ina} 
  A.~A.~H.~Graham and R.~Jha,
  Phys.\ Rev.\ D {\bf 90}, no. 4, 041501 (2014)
  [arXiv:1407.6573 [gr-qc]].

  
\bibitem{Herdeiro:2014goa} 
  C.~A.~R.~Herdeiro and E.~Radu,
  Phys.\ Rev.\ Lett.\  {\bf 112}, 221101 (2014)
  [arXiv:1403.2757 [gr-qc]].
  
\bibitem{Damour:1993hw} 
  T.~Damour and G.~Esposito-Farese,
  Phys.\ Rev.\ Lett.\  {\bf 70}, 2220 (1993).
  
\bibitem{Damour:1996ke} 
  T.~Damour and G.~Esposito-Farese,
  Phys.\ Rev.\ D {\bf 54}, 1474 (1996)
  [gr-qc/9602056].
  
\bibitem{Cardoso:2013fwa} 
  V.~Cardoso, I.~P.~Carucci, P.~Pani and T.~P.~Sotiriou,
  Phys.\ Rev.\ Lett.\  {\bf 111}, 111101 (2013)
  [arXiv:1308.6587 [gr-qc]].
  
\bibitem{Cardoso:2013opa} 
  V.~Cardoso, I.~P.~Carucci, P.~Pani and T.~P.~Sotiriou,
  Phys.\ Rev.\ D {\bf 88}, 044056 (2013)
  [arXiv:1305.6936 [gr-qc]].

\bibitem{Kanti:1995vq} 
  P.~Kanti, N.~E.~Mavromatos, J.~Rizos, K.~Tamvakis and E.~Winstanley,
  Phys.\ Rev.\ D {\bf 54}, 5049 (1996)
  [hep-th/9511071].
  
\bibitem{Sotiriou:2013qea} 
  T.~P.~Sotiriou and S.~Y.~Zhou,
  Phys.\ Rev.\ Lett.\  {\bf 112}, 251102 (2014)
  [arXiv:1312.3622 [gr-qc]].
  
\bibitem{Sotiriou:2014pfa} 
  T.~P.~Sotiriou and S.~Y.~Zhou,
  Phys.\ Rev.\ D {\bf 90}, no. 12, 124063 (2014)
  [arXiv:1408.1698 [gr-qc]].
  
\bibitem{Mignemi:1992pm} 
  S.~Mignemi and N.~R.~Stewart,
  Phys.\ Lett.\ B {\bf 298}, 299 (1993)
  [hep-th/9206018].
  
\bibitem{Yunes:2011we} 
  N.~Yunes and L.~C.~Stein,
  Phys.\ Rev.\ D {\bf 83}, 104002 (2011)
  [arXiv:1101.2921 [gr-qc]].

  
\bibitem{Horndeski:1974wa} 
  G.~W.~Horndeski,
  Int.\ J.\ Theor.\ Phys.\  {\bf 10}, 363 (1974).
  
\bibitem{Deffayet:2011gz} 
  C.~Deffayet, X.~Gao, D.~A.~Steer and G.~Zahariade,
  Phys.\ Rev.\ D {\bf 84}, 064039 (2011)
  [arXiv:1103.3260 [hep-th]].
  
\bibitem{Nicolis:2008in} 
  A.~Nicolis, R.~Rattazzi and E.~Trincherini,
  Phys.\ Rev.\ D {\bf 79}, 064036 (2009)
  [arXiv:0811.2197 [hep-th]].
  
\bibitem{Deffayet:2013lga} 
  CŽ.~Deffayet and D.~l.~A.~Steer,
  Class.\ Quant.\ Grav.\  {\bf 30}, 214006 (2013)
  [arXiv:1307.2450 [hep-th]].
  
\bibitem{Hui:2012qt} 
  L.~Hui and A.~Nicolis,
  Phys.\ Rev.\ Lett.\  {\bf 110}, 241104 (2013)
  [arXiv:1202.1296 [hep-th]].
  
\bibitem{Kobayashi:2011nu} 
  T.~Kobayashi, M.~Yamaguchi and J.~'i.~Yokoyama,
  Prog.\ Theor.\ Phys.\  {\bf 126}, 511 (2011)
  [arXiv:1105.5723 [hep-th]].
  
\bibitem{Babichev:2013cya} 
  E.~Babichev and C.~Charmousis,
  JHEP {\bf 1408}, 106 (2014)
  [arXiv:1312.3204 [gr-qc]].
  
\bibitem{hartle67} 
  J.~B.~Hartle,
  Astrophys.\ J.\  {\bf 150}, 1005 (1967).
  
\bibitem{Horava:2009uw} 
  P.~Horava,
  Phys.\ Rev.\ D {\bf 79}, 084008 (2009)
  [arXiv:0901.3775 [hep-th]].
  
\bibitem{Blas:2009qj} 
  D.~Blas, O.~Pujolas and S.~Sibiryakov,
  Phys.\ Rev.\ Lett.\  {\bf 104}, 181302 (2010)
  [arXiv:0909.3525 [hep-th]].
  
\bibitem{Jacobson:2010mx} 
  T.~Jacobson,
  Phys.\ Rev.\ D {\bf 81}, 101502 (2010)
  [Phys.\ Rev.\ D {\bf 82}, 129901 (2010)]
  [arXiv:1001.4823 [hep-th]].
  
\bibitem{Barausse:2011pu} 
  E.~Barausse, T.~Jacobson and T.~P.~Sotiriou,
  Phys.\ Rev.\ D {\bf 83}, 124043 (2011)
  [arXiv:1104.2889 [gr-qc]].
  
\bibitem{Barausse:2013nwa} 
  E.~Barausse and T.~P.~Sotiriou,
  Class.\ Quant.\ Grav.\  {\bf 30}, 244010 (2013)
  [arXiv:1307.3359 [gr-qc]].
  
\bibitem{Blas:2011ni} 
  D.~Blas and S.~Sibiryakov,
  Phys.\ Rev.\ D {\bf 84}, 124043 (2011)
  [arXiv:1110.2195 [hep-th]].
  
\bibitem{Barausse:2012ny} 
  E.~Barausse and T.~P.~Sotiriou,
  Phys.\ Rev.\ Lett.\  {\bf 109}, 181101 (2012)
  [Phys.\ Rev.\ Lett.\  {\bf 110}, no. 3, 039902 (2013)]
  [arXiv:1207.6370].
  
\bibitem{Barausse:2012qh} 
  E.~Barausse and T.~P.~Sotiriou,
  Phys.\ Rev.\ D {\bf 87}, 087504 (2013)
  [arXiv:1212.1334].
  
\bibitem{Berglund:2012bu} 
  P.~Berglund, J.~Bhattacharyya and D.~Mattingly,
  Phys.\ Rev.\ D {\bf 85}, 124019 (2012)
  [arXiv:1202.4497 [hep-th]].
  
\bibitem{Sotiriou:2014gna} 
  T.~P.~Sotiriou, I.~Vega and D.~Vernieri,
  Phys.\ Rev.\ D {\bf 90}, no. 4, 044046 (2014)
  [arXiv:1405.3715 [gr-qc]].
  
\bibitem{Bhattacharyya:2014kta} 
  J.~Bhattacharyya and D.~Mattingly,
  Int.\ J.\ Mod.\ Phys.\ D {\bf 23}, no. 13, 1443005 (2014)
  [arXiv:1408.6479 [hep-th]].

%
%
%
%
.



\end{thebibliography}
\end{document}